\documentclass[conference, 10pt]{IEEEtran}

\usepackage{textcomp}
\usepackage{xcolor}
\usepackage{import}
\usepackage{hyperref}
\usepackage{tabularx}
\usepackage{multirow}
\usepackage{algorithm}
\usepackage[noend]{algpseudocode}
\usepackage{amsmath}
\usepackage{graphicx}
\usepackage{subcaption}
\usepackage{mathtools}
\usepackage{booktabs} 
\usepackage{amssymb}
\usepackage{supertabular}
\usepackage{float}
\usepackage{flushend}

\usepackage{tikz}

\AtBeginDocument{%
  }

\newcommand{\concat}{\mathbin{\|}}
\newcommand{\abs}[1]{\lvert #1 \rvert}

\begin{document}

\title{Secure Multi-Path Routing with All-or-Nothing Transform for Network-on-Chip Architectures}

% \title{\huge All-or-nothing and multi-path routing Network-onChips}

%\thanks{This work was partially supported by NSF grant SaTC-1936040}

\author{\IEEEauthorblockN{Hansika Weerasena, Matthew Randall and Prabhat Mishra}
\IEEEauthorblockA{Department of Computer \& Information Science \& Engineering\\
University of Florida, Gainesville, Florida, USA}}
% \author{Hansika~Weerasena,~\IEEEmembership{Student Member,~IEEE,}
% and~Prabhat~Mishra,~\IEEEmembership{Fellow,~IEEE,}% <-this % stops a space
% \IEEEcompsocitemizethanks{\IEEEcompsocthanksitem H. Weerasena, and P. Mishra are with the Department of Computer \& Information Science \& Engineering, University of Florida, Gainesville, Florida, USA.}
% }

\maketitle

\begin{abstract}

Ensuring Network-on-Chip (NoC) security is crucial to design trustworthy NoC-based System-on-Chip (SoC) architectures. While there are various threats that exploit on-chip communication vulnerabilities, eavesdropping attacks via malicious nodes are among the most common and stealthy. Although encryption can secure packets for confidentiality, it may introduce unacceptable overhead for resource-constrained SoCs. In this paper, we propose a lightweight confidentiality-preserving framework that utilizes a quasi-group based All-Or-Nothing Transform (AONT) combined with secure multi-path routing in NoC-based SoCs. By applying AONT to each packet and distributing its transformed blocks across multiple non-overlapping routes, we ensure that no intermediate router can reconstruct the original data without all blocks. Extensive experimental evaluation demonstrates that our method effectively mitigates eavesdropping attacks by malicious routers with negligible area and performance overhead. Our results also reveal that AONT-based multi-path routing can provide 7.3x reduction in overhead compared to traditional encryption for securing against eavesdropping attacks.

\end{abstract}

% \begin{IEEEkeywords}
% network-on-chip, distributed-denial-of-service, on-chip communication security, graph neural networks.
% \end{IEEEkeywords}

\pagestyle{empty}

\section{Introduction}
\label{sec:introduction}

%Parallel workloads and domain-specific accelerators, such as neural network engines, are reshaping computing performance and architectural design. 
Modern heterogeneous Systems-on-Chip (SoCs) and Multi-Processor SoCs (MPSoCs) now integrate hundreds of Intellectual Property (IP) cores to meet the growing application demands. Network-on-Chip (NoC) enables scalable, low-latency, high-bandwidth and efficient communication among these cores through an interconnected set of routers. Since the NoC facilitates communication between all IP cores, it presents an ideal threat vector-a goldmine for attackers to exploit and pose a significant security challenge. The integration of third-party IPs due to market timing constraints and reliance on untrusted foundries increase the risk of hardware Trojans and compromised routers, which can undermine system confidentiality, integrity, or availability.

\begin{figure}[tp]
\centering
\vspace{-0.1in}
\includegraphics[width=0.80\columnwidth]{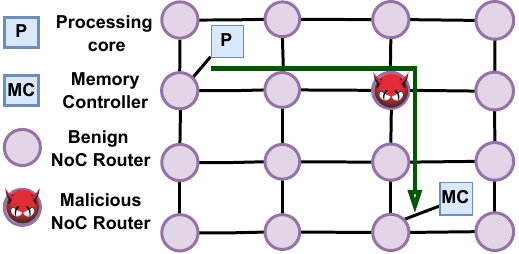}
\vspace{-0.05in}
\caption{An example of eavesdropping attack in a 4x4 mesh NoC. A malicious router can eavesdrop on communications between a processing core (P) an a directory controller (DC).}
\label{fig:into}
\vspace{-0.2in}
\end{figure}

NoC security vulnerabilities across different security goals have been extensively studied~\cite{weerasena2024security}. Figure~\ref{fig:into} shows an illustrative example of eavesdropping attack, which is one of the most common and stealthy threats to on-chip communication. It is a passive attack where a compromised router silently monitors flits as they traverse the network, potentially leaking sensitive information such as encryption keys, inference results, or memory addresses. Traditional defenses, including cryptographic encryption (e.g., AES-CTR), provide strong confidentiality guarantees but may introduce significant area, performance, and energy overhead, making them impractical for many SoC designs. These limitations highlight the need for a lightweight solution that can provide strong protection against passive threats like eavesdropping.  We propose a lightweight confidentiality-preserving framework for on-chip communication, making it well-suited for resource-constrained SoCs. This paper makes the following contributions.

\begin{itemize}
    \item We develop a lightweight All-Or-Nothing Transform (AONT) that splits a message into multiple blocks, all of which are needed to reconstruct the original.

    \item We propose a non-overlapping multipath routing strategy for mesh NoCs to distribute transformed blocks.
    
    \item We evaluate our approach across diverse scenarios, demonstrating strong confidentiality with minimal area and performance overhead.

\end{itemize}

While chaffing and winnowing–based AONT has been used for securing NoC~\cite{weerasena2021lightweight}, our work introduces a synergetic integration of AONT with disjoint multipath routing, eliminating redundant traffic and achieving superior performance.

\begin{figure*}[tb]
\vspace{-0.2in}
\centering
\includegraphics[width=0.95\textwidth]{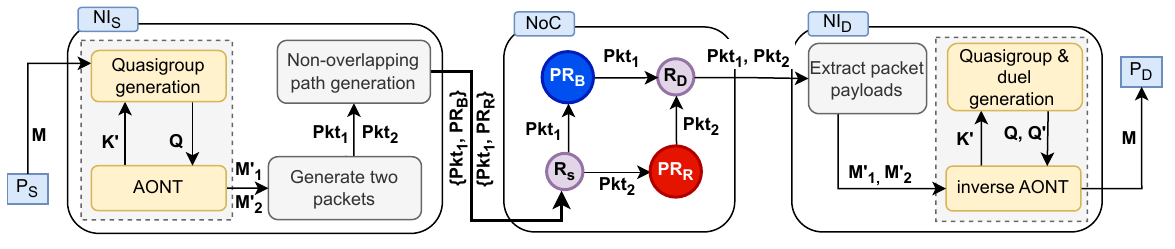}
\vspace{-0.15in}
\caption{Overview of the proposed all-or-nothing transform (AONT)-based multi-path routing framework.}
\label{fig:overview}
\vspace{-0.2in}
\end{figure*}

% \import{sections/}{background_and_related_work.tex}
\section{Threat Model}
\label{sec:threat}

In NoC-based MPSoCs, the NoC plays a key role in communicating cache coherence messages for directory-based protocols, such as MESI or MOESI. On a local cache miss, the core issues a control packet (e.g., load or store request) to the shared L2 or last-level cache (LLC), which may forward the request to memory if the data is not cached. Once the data is located, either from the LLC or main memory, a response is sent back to the requesting core as a data packet. This round-trip communication takes place entirely over the NoC, often passing through multiple intermediate routers. Here, we assume a widely adopted mesh NoC topology. Figure~\ref{fig:into} illustrates an eavesdropping attack in a 4x4 mesh NoC, where a single malicious router is strategically placed to eavesdrop. A single malicious router is widely used as a realistic threat model since it would be hard to detect during design time due to its minimal area and power footprint. The malicious router can intercept in-transit data packets between legitimate nodes. This allows it to extract sensitive information, such as cryptographic keys, proprietary deep learning model parameters, or private user data originating from memory, posing a serious confidentiality threat in safety- and security-critical systems.
\section{Secure multi-path routing with AONT}

% \begin{figure}[tp]
% \centering
% \vspace{-0.1in}
% \includegraphics[width=0.90\columnwidth]{./figures/multi_aont_NoC.pdf}
% \vspace{-0.1in}
% \caption{AONT with multi-path routing. The malicious node cannot get all AONT blocks.}
% \label{fig:into}
% \vspace{-0.1in}
% \end{figure}

% To enhance confidentiality in on-chip communication, we implement a lightweight multi-path routing with AONT mechanism in mesh NoCs. There are two main components of our approch (1). Quasigroup-based All-orNothing Transform and (2). Non-overlapping multipath routing. Figure~\ref{fig:overview} shows an overview of our approach, when the payload $M$ needs to be sent from $P_S$ to $P_D$. The source network interface ($NI_S$) will first generate a quasigroup ($Q$) using the pre-shared key $K'$ between the source and destination. AONT transform will occur using $Q$ and create a pseudo-message $M'$. Then the $M'$ will be divided into two portions and generate two packets as $Pkt_1$ and $Pkt_2$. The non-overlapping path generation will create two piovt routers ($PR_1$ and $PR_2$). First, $Pkt_1$ will go to $PR_1$ as pivot router and $Pkt_2$ go to $PR_2$ and then they will be routed to actual destination router ($R_D$). This approach guarantees that a single malicious router cannot recive all the AONT blocks, which prevent it form leaking even partial information. Finally the source router will merge two packet payloads and recreate pesudo-message ($M'$). Finally, inverse AONT transform will reconstruct actual message $M$. The rest of the section will describe this methadlaogy in detail.  

To enhance confidentiality in on-chip communication, we implement a lightweight multi-path routing mechanism combined with an All-Or-Nothing Transform (AONT) in mesh NoCs. Our approach consists of two main components: (1) a quasigroup-based AONT and (2) non-overlapping multi-path routing. Figure~\ref{fig:overview} illustrates the process when a payload $M$ is sent from source $P_S$ to destination $P_D$. The source network interface ($NI_S$) first generates a quasigroup $Q$. The AONT uses $Q$ to transform $M$ into two pseudo-messages, $M'_1$ and $M'_2$, which serve as the payloads for packets $Pkt_1$ and $Pkt_2$. Non-overlapping path generation selects two pivot routers ($PR_B$ and $PR_R$); $Pkt_1$ is routed via $PR_B$ and $Pkt_2$ via $PR_R$, before both reach the destination router ($R_D$). Non-overlapping paths ensure that no single malicious router can access all AONT blocks, thereby guaranteeing complete confidentiality with no possibility of even partial data leakage. At the destination, the network interface extracts the payloads of the two packets, $M'_1$ and $M'_2$. Finally, the inverse AONT uses quasigroup and the dual to recover the original message $M$. The rest of this section describes three important steps in our methodology: AONT and packetization, multi-path routing, and packet reassembly and inverse AONT.

% \textcolor{blue}{only data packets being enc, option 2 added padding and still perform better than AES}

\subsection{All or Nothing Transformation and Packetization} 
\label{CnW:subSec:methodalogy-all_or_nothing_transform}

\begin{algorithm}
  \caption{AONT and Packetization}
  \label{CnW:alg:All_or_nothing_trasnformation}
  \begin{algorithmic}[1]
    \Function{AONT\_and\_Packet}{$M$}
      \State $M$ $\gets$ $B_1 \concat ... \concat B_{s}$ where $B_i = (h_{i1},...,h_{in} )$
      \State $K' \gets$ random permutation
      \State $\langle Q, \bullet \rangle$, \_ $\gets Quasi(K')$
      \State $l_1 \leftarrow k_1, l_2 \leftarrow k_2 \bullet l_1,...,l_n \leftarrow k_n \bullet l_{n-1}$ and $l \leftarrow l_n$
      \For{$i= 1$ to $s$} \label{line:alg3:ref1}
        \State $I(i) \gets $ base-$n$ representation of $i$ with $\abs{I(i)} = n$
        \State $R(i) \gets (r_{i1},...,r_{in} ) $ where $r_{in} \gets l \bullet i_{in}, r_{in-1} \gets r_{in} \bullet i_{in-1},..., r_{i1} \gets r_{i2} \bullet i_{i1}$ 
        \State $B'_i \gets (h'_{i1},...,h'_{in} )$ where $h'_{ij} \leftarrow r_{ij} \bullet h_{ij}$
      \EndFor 
      \State $C_1 \leftarrow B'_1$
      \For{$i = 2$ to $s$}
        \State $C_i \gets C_{i-1} * B_i$
      \EndFor
      \State $B'_{s+1} \gets C_s*K'$
      \State $M'_1$, $M'_2 \gets B'_1 \concat ... \concat B'_{s/2}$,  
 $B'_{s/2 + 1} \concat ... \concat B'_{s+1}$
      \State $Pkt_1$, $Pkt_2 \gets$ \textit{head($Pkt_1$)}$\concat M'_1$,    \textit{head($Pkt_2$)} $\concat M'_2$
      \State \textbf{return} $Pkt_1$, $Pkt_2$
    \EndFunction
  \end{algorithmic}
\end{algorithm}

AONT is a cryptographic primitive that transforms a message into multiple blocks such that all blocks are needed to recover the original message, any missing block prevents even partial message recovery. Algorithm~\ref{CnW:alg:All_or_nothing_trasnformation} describes the AONT using quasigroups and packet generation. The message is reformatted to support arithmetic operations in base~$n$ (line~2). Here, the message is divided into $s$ blocks $B_1, B_2, \dots, B_s$, where each block $B_i = (h_{i1}, h_{i2}, \dots, h_{in})$ contains $n$ elements. Each element $h_{ij}$ is represented in base~$n$ using $\log_2 n$ bits. For instance, when $n = 4$, $(01)_2$ maps to 1 and $(11)_2$ to 3. For less computationally demanding base conversion, $n$ is chosen as a power of two, which motivates using a Fermat prime for $p$. Fermat primes, of the form $2^{2^z} + 1$, where $z$ is an integer, enables efficient modular arithmetic with base-$n$ values. The call to Algorithm~\ref{CnW:alg:generate_quasi_group_and_dual} constructs the quasigroup $\langle Q, \bullet \rangle$ using the randomly generated key $K'$ (line 4). A finite quasigroup of order $n$ defines a binary operation $\bullet$ over a set $Q$ such that, for any $a, b \in Q$, the equations $a \bullet x = b$ and $y \bullet a = b$ have unique solutions for $x$ and $y$. The \textit{leader} $l$ is derived by iteratively applying $\bullet$ over elements of $K'$ (line 5). 

Lines 6–9 transform each message block $B_i$ into a pseudo-message block $B'_i$ as follows: line~7 represents the integer $i$ in base-$n$ as a length-$n$ vector $I(i)$ (e.g., $i = 1 = (0001)_4$ in base-4 becomes $(4,4,4,1)$, where the character 4 denotes 0). In line~8, an intermediate block $R(i)$ is computed by applying the binary operator $\bullet$ to the elements of $I(i)$, using the leader $l$ as the $n^{\text{th}}$ element. Finally, line~9 produces the pseudo-block $B'_i$ by element-wise application of $\bullet$ between $B_i$ and $E(i)$. Lines~10–13 compute the final pseudo-block $B'_{s+1}$ by iteratively combining blocks $C_i = C_{i-1} * B_i$, where $C_1 = B'_1$, and $*$ is defined as $(a_i \cdot b_i) \mod p$ for each element. Then, the pseudo message $M'_1$ is formed by concatenating $B'_i$ to $B'_{s/2}$ and $M'_2$ by concatnating the rest of AONT blocks (line~14). Finally, $Pkt_1$ is created by using $M'_1$ as payload and $Pkt_2$ is created by using $M'_2$ as payload (line~15). \textit{head()} function will generate headers for respective packets.

% \begin{itemize}
%     \item Line~6: Integer $i$ is represented in base-$n$ as a length-$n$ vector $I(i)$ (e.g., $i=1 = (0001)_4$ in base-4 becomes $(4,4,4,1)$ as charter 4 represent 0.
%     \item Line~7: Intermediate block $E(i)$ is computed using the $\bullet$ operator over elements of $I(i)$, while using leader $l$ as $n^{th}$ element.
%     \item Line~8: Pseudo-block $B'_i$ is generated by element-wise application of $\bullet$ between $B_i$ and $E(i)$.
% \end{itemize}

\subsubsection{Quasigroups Generation}
\label{CnW:subSec:methodalogy-Quasigroups_and_latin_squares} 

Our approach adopts the quasigroup generation technique proposed by Marnes et al.~\cite{marnas2003all}, as it enables a fast and randomized construction process (Algorithm~\ref{CnW:alg:generate_quasi_group_and_dual}). A Latin Square (LS) is an $n \times n$ matrix over $n$ symbols, where each symbol appears exactly once in every row and column. The multiplication table of a quasigroup of order $n$ is a Latin Square of the same size. In our case, $n = p - 1$, where $p$ is a prime. The first row of the LS is initialized as a random permutation and provided as input to the algorithm (line~2). Each element in the $i$-th row, for $i = 2, \dots, n$, is computed using the formula $(i \times q_{1j}) \bmod p$ (line~5). The resulting LS is used in the AONT transformation. Algorithm~\ref{CnW:alg:generate_quasi_group_and_dual} also generates the dual of the quasigroup simultaneously, although only the inverse AONT at the receiver side requires it. For each element $q_{ij}$ in row $i$ and column $j$, the corresponding dual element $q'_{ix}$ is set to $j$, where $x = q_{ij}$ (lines~6 and 7).

\begin{algorithm}
  \caption{Generating Quasigroup with Dual}
  \label{CnW:alg:generate_quasi_group_and_dual}
  \begin{algorithmic}[1]
    \Function{$Quasi$}{$K'$}
      \State $q_{11} \concat q_{12} \concat ... \concat q_{1n}$ $\gets K'$ 
      \For{$i= 1$ to $n$}
        \For{$j= 1$ to $n$} \label{CnW:alg_line:generate_quasi_group_and_dual-line4}
            \State $q_{ij} \leftarrow (i \times q_{1j}) \mod{p}$
            \State $x \leftarrow q_{ij}$ \label{CnW:alg_line:generate_quasi_group_and_dual-line6}
            \State $q'_{ix} \leftarrow j$ \label{CnW:alg_line:generate_quasi_group_and_dual-line7}
        \EndFor
      \EndFor
      \State $\langle Q, \bullet \rangle$, $\langle Q, \circ \rangle \gets$ LS of $q_{ij}$, LS of $q'_{ix}$
      % \State $\langle Q, \circ \rangle \gets$ LS of $q'_{iz}$
      \State \textbf{return} $\langle Q, \bullet \rangle$, $\langle Q, \circ \rangle$
    \EndFunction
  \end{algorithmic}
\end{algorithm}

\vspace{-0.1in}
\subsection{Non-overlapping Path Generation}

\begin{figure}[htp]
\centering
\vspace{-0.15in}
\includegraphics[width=0.8\columnwidth]{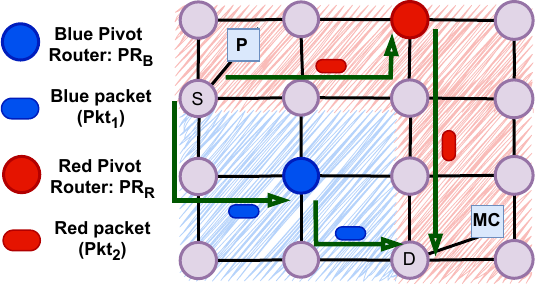}
 \vspace{-0.1in}
\caption{NoC is divided into two regions: blue and red}
\label{fig:non_overlapping}
\vspace{-0.2in}
\end{figure}

$Pkt_1$ and $Pkt_2$ are routed along two disjoint paths to ensure that no single router can intercept all AONT blocks. NoC is divided into two regions, blue and red, based on the source's position relative to the destination. The blue region is selected according to the rules in Table~\ref{tab:blue_region_selection}; the remaining area forms the red region. Figure~\ref{fig:non_overlapping} illustrates the case when the destination lies bottom-right of the source. Red region uses XY routing (X first, then Y) and blue uses YX routing (Y first, then X). Note that virtual channels are equally divided between XY and YX to avoid deadlocks.  A random pivot router ($PR$) is selected from each region. The source sends one part of the message to the red $PR$ and the other to the blue $PR$, and both then forward their parts to the actual destination. To support two-hop routing via intermediate nodes, each packet ($Pkt_1$ and $Pkt_2$) includes the pivot router and the final destination ($fin\_id$) as metadata in packet header. At the pivot router, the packet header is updated to route to the final destination.

\begin{table}[ht]
\centering
\vspace{-0.05in}
\caption{Blue Region Selection Rules}
\vspace{-0.1in}
\begin{tabular}{|l|l|}
\hline
\textbf{Case/Scenario} & \textbf{Blue Region Square} \\
\hline
\texttt{dest} is top-right of \texttt{src} & Above \texttt{src} and left of \texttt{dest} \\
\hline
\texttt{dest} is bottom-right of \texttt{src} & Below \texttt{src} and left of \texttt{dest} \\
\hline
\texttt{dest} is top-left of \texttt{src} & Above \texttt{src} and right of \texttt{dest} \\
\hline
\texttt{dest} is bottom-left of \texttt{src} & Below \texttt{src} and right of \texttt{dest} \\
\hline
\end{tabular}
\label{tab:blue_region_selection}
\vspace{-0.1in}
\end{table}

Two special cases arise when the source and destination lie on the same row or column. In such cases, the NoC is divided into top-bottom or left-right halves, respectively, and each packet is routed through a pivot router in a different half. However, using the default routing strategy may lead to path overlaps in these scenarios. To address this, we introduce an optional \textit{flip\_route} flag that allows the blue path to flip its routing strategy (from YX to XY) at the blue pivot router. This flip is triggered when the source and destination lie on the same row or column, where spatial constraints can cause potential path overlap. Flipping the blue path’s routing strategy ensures disjoint routes in these two special cases. This method guarantees spatial separation of the two paths, ensuring confidentiality against a eavesdropping malicious router. 

\subsection{Packet Reassembly and Inverse AONT}

Algorithm~\ref{CnW:alg:All_or_nothing_trasnformation_inverse} outlines the packet reassembly and inverse $AONT$ process. At the destination node, the payloads of $Pkt_1$ and $Pkt_2$ are first reassembled based on their packet sequence numbers and their payload is extracted (line 2). Once combined, the inverse AONT is applied to reconstruct the original message with the following key differences from the \textit{AONT} process. First, in line~6, the key $K'$ is recovered via element-wise division of $C_s$ by $B'_{s+1}$ using the relation $(a \cdot k') \bmod p = m' \Longleftrightarrow a / m' = k'$. Then, in line~7, the quasigroup along with its corresponding dual are reconstructed. Finally, in line~12, the original message blocks are recovered from the pseudo blocks using the dual operator $\circ$ of $\bullet$. Here, a dual binary operator $\circ$ can be defined for a given binary operator $\bullet$ over the set $Q$ such that $a \circ b = c$ if and only if $a \bullet c = b$. This relationship defines a new finite quasigroup $\langle Q, \circ \rangle$, which is the dual of $\langle Q, \bullet \rangle$. The following identities hold: $a \circ (a \bullet b) = b$ and $a \bullet (a \circ b) = b$.

\begin{algorithm}
  \caption{Packet reassembly and Inverse AONT}
   \label{CnW:alg:All_or_nothing_trasnformation_inverse}
  \begin{algorithmic}[1]
    \Function{Reassemble\_and\_$AONT^{-1}$}{$Pkt_1,Pkt_2$}
      \State $M'_1$, $M'_2$ $\gets$ extract payload from $Pkt_1$ and $Pkt_2$
      \State $B'_1 \concat ... \concat B'_{s}$ $\gets M'_1 \concat M'_2$ where $B'_i = (h'_{i1},...,h'_{in} )$
      \State $C_1 \gets B'_1$
      \For{$i=2$ to $s$} \label{line:alg4:ref1}
        \State $C_i \gets C_{i-1}*B'_i$
      \EndFor
      \State $K'\gets C_s/B'_{s+1}$
      \label{CnW:alg_line:All_or_nothing_trasnformation_inverse-line8}
      \State $\langle Q, \bullet \rangle$, $\langle Q, \circ \rangle \gets Quasi(K')$
       \label{CnW:alg_line:All_or_nothing_trasnformation_inverse-line9}
      \State $l_1 \gets k_1, l_2 \gets k_2 \bullet l_1,...,l_n \gets k_n \bullet l_{n-1}$ and $l \gets l_n$
      \For{$i= 1$ to $s$} \label{line:alg4:ref2}
        \State $I(i) \gets $ base-$n$ representation of $i$ with $\abs{I(i)} = n$
        \State $R(i) \gets (r_{i1},...,r_{in} ) $ where $e_{in} \gets l \bullet i_{in}, r_{in-1} \gets r_{in} \bullet i_{in-1},..., r_{i1} \gets r_{i2} \bullet i_{i1}$ 
        \State $B(i) \gets (h_{i1},...,h_{in} )$ where $h_{ij} \gets r_{ij} \circ h'_{ij}$
        \label{CnW:alg_line:All_or_nothing_trasnformation_inverse-line14}
      \EndFor
      \State $M \gets B_i \concat B_2 \concat ... \concat B_{s+1}$
      \State \textbf{return} $M$
    \EndFunction
  \end{algorithmic}
\end{algorithm}

\section{Experiments}
\label{sec:experiments}

% \begin{table}[tbp]
% % \vspace{-0.1in}
% \caption{System and interconnect configurations.}
% \vspace{-0.15in}
% \label{tab:configTable}
% \begin{center}
% \begin{tabular}{|p{0.50\columnwidth} | p{0.40\columnwidth}|}
% \hline
% \textbf{Parameter} & \textbf{Details} \\
% \hline
% Topology & 8x8 mesh with 64 nodes \\
% \hline
% Processor configurations & X86, 2GHz \\
% \hline
% Cache coherency protocol & MESI two-level\\
% \hline
% L1 instruction \& data cache & 32KB, 32KB (private) \\
% \hline
% L2 cache & 512KB (shared) \\
% \hline
% \end{tabular}
% \label{tab1}
% \end{center}
% \vspace{-0.2in}
% \end{table}

% To evaluate the effectiveness of our approach, we simulated in 4x4 and 8x8 MPSoC. We used both the cycle-accurate gem5~\cite{gem5} and Noxim~\cite{catania2016cycle} simulators. Full system simulations were performed in gem5 to collect network traces, utilizing seven benchmarks from SPLASH-2~\cite{woo1995splash} and PARSEC~\cite{bienia2008parsec}: \textit{fft}, \textit{fmm}, \textit{lu}, \textit{barnes}, \textit{radix}, \textit{blackscholes}, and \textit{ocean}. We modified noxim with AONT and inverse AONt at NIs and multipath selection at NI and routers to support the handvover to actual destination from pivot node in the network interface (processing element in noxim) and  (delays added to simulate hardware and encryption). Since our AONT implementation works on blocks in parallel, we compared it with AES-128 in parallel CTR mode of encryption to enable a fair comparison. So we compare with 3 cases: No Security, Our Method and AES-128 CTR mode or similar.

% \subsection{Experimental Setup}

We used a combined setup with cycle-accurate gem5~\cite{gem5} and Noxim~\cite{catania2016cycle}. Full-system simulations in gem5 generated network traces using seven SPLASH-2~\cite{woo1995splash} and PARSEC~\cite{bienia2008parsec} benchmarks, which were then used for trace-driven simulations in Noxim. The simulated system features a two-level MESI cache coherence protocol, x86 cores operating at 2GHz, and an 8×8 mesh interconnect with 64 nodes. Each core has 32KB private L1 I/D caches, and all cores share a 512KB L2 cache. We extended Noxim to support our proposed security mechanism by integrating AONT, inverse AONT and multi-path route generation at the NIs, and destination swap support at routers. We compare against AES-128, a widely adopted standard, as lightweight ciphers like Hummingbird lack standardization and have known vulnerabilities. Since our design supports parallel AONT transformation, we compare it against AES-128 in parallel CTR mode to ensure a fair comparisson. The evaluation includes three scenarios: (1) No Security, (2) Our Proposed Method, and (3) AES-128 in CTR mode.

\subsection{Experimental Results}

Table~\ref{tab:res1} presents the average delay across seven benchmarks on an 8×8 NoC for three configurations: no security, AES-128, and our proposed method. For each benchmark, we conducted experiments over 20 random application-to-core mappings, and the reported values reflect the average delay across these runs. Compared to the no-security baseline (12.9 cycles), AES-128 increases average delay by over 20x (261.5 cycles), while our method incurs only a 2.8x increase (35.9 cycles). This demonstrates that our AONT-based approach provides strong confidentiality with significant (7.3x) reduction in performance overhead compared to traditional encryption. We also compared our approach with the chaffing and winnowing-based AONT scheme from~\cite{weerasena2021lightweight}; our approach achieves superior performance with 83.9\% reduction in average packet latency. This is because the chaffing process significantly increase the packet sizes ($\approx 4x$) due to multiple counters and authentication tags.

\begin{table}[ht]
\centering
\caption{Performance comparison for 8x8 mesh NoC}
\vspace{-0.05in}
\label{tab:res1}
\begin{tabular}{|c|c|c|c|c|c|c|}
\hline
\multicolumn{1}{|c|}{\textbf{Benchmark}} & \multicolumn{2}{c|}{\textbf{\begin{tabular}[c]{@{}l@{}} Delay: No \\ Security \end{tabular}}} & \multicolumn{2}{c|}{\textbf{\begin{tabular}[c]{@{}l@{}} Delay: \\ AES-128  \end{tabular}}} & \multicolumn{2}{c|}{\textbf{\begin{tabular}[c]{@{}l@{}} Delay: \\ Proposed \end{tabular}}} \\
\cline{2-7}
\multicolumn{1}{|c|}{ } & \textbf{Avg } & \textbf{Max } & \textbf{Avg } & \textbf{Max } & \textbf{Avg } & \textbf{Max } \\
\hline 
FFT & 13.2 & 31 & 259.1 & 630 & 36.5 & 97 \\
OCEAN & 13.1 & 33 & 258.3 & 626 & 35.1 & 95 \\
RADIX & 14.1 & 33 & 259.5 & 628 & 34.2 & 99 \\
FMM & 10.9 & 33 & 266.9 & 630 & 35.5 & 101 \\
LU & 13.6 & 31 & 258.2 & 628 & 37.0 & 99 \\
BARNES & 11.2 & 31 & 270.5 & 630 & 36.2 & 101 \\
black & 13.9 & 31 & 257.9 & 626 & 36.6 & 99 \\
\hline
Average & 12.9 & 33 & 261.5 & 630 & 35.9 & 101 \\
\hline
\end{tabular}
\vspace{-0.2in}
\end{table}

\begin{table}[ht]
\centering
\caption{Performance comparison for 8x8: 2 benchmarks}
\vspace{-0.05in}
\label{tab:res2}
\begin{tabular}{|c|c|c|c|c|c|c|}
\hline
\multicolumn{1}{|c|}{\textbf{Benchmark}} & \multicolumn{2}{c|}{\textbf{\begin{tabular}[c]{@{}l@{}} Delay: No \\ Security \end{tabular}}} & \multicolumn{2}{c|}{\textbf{\begin{tabular}[c]{@{}l@{}} Delay: \\ AES-128  \end{tabular}}} & \multicolumn{2}{c|}{\textbf{\begin{tabular}[c]{@{}l@{}} Delay: \\ Proposed \end{tabular}}} \\
\cline{2-7}
\multicolumn{1}{|c|}{ } & \textbf{Avg } & \textbf{Max } & \textbf{Avg } & \textbf{Max } & \textbf{Avg } & \textbf{Max } \\
\hline 
FFT  + 1 & 14.1 & 60 & 263.9 & 658 & 40.2 & 95 \\
OCEAN  + 1 & 14.9 & 50 & 261.7 & 654 & 39.9 & 95 \\
RADIX  + 1 & 15.1 & 57 & 263.1 & 640 & 38.7 & 97 \\
FMM  + 1 & 11.3 & 57 & 272.4 & 652 & 39.0 & 91 \\
LU  + 1 & 14.3 & 50 & 262.6 & 654 & 39.4 & 95 \\
BARNES  + 1 & 12.2 & 56 & 272.2 & 665 & 39.2 & 95 \\
black  + 1 & 14.7 & 54 & 265.3 & 654 & 40.1 & 105 \\
\hline
Average & 13.8 & 60 & 266.5 & 665 & 39.5 & 105 \\
\hline % Bottom line
\end{tabular}
\end{table}

Table~\ref{tab:res2} shows the average delay for experiments where two benchmarks were run concurrently on an 8×8 NoC. Similarly to the single-benchmark case, we averaged results over 20 random application mappings. If we focus on table entry (e.g., \textit{FFT + 1}), each main benchmark (e.g., \textit{FFT}) is paired with a second benchmark chosen randomly in each of the 20 runs. Compared to the no-security baseline (13.8 cycles), AES-128 increases average delay by over 19× (266.5 cycles), while our proposed method results in only a 2.9× increase (39.5 cycles), confirming its efficiency under multi-application workloads.

\color{black}

% \color{red}

% \textbf{Next Steps:}

% \begin{itemize}
%     \item  Run non-modified code and check if the same issue is happening (generating traffic not in the table).
%     \item Fix the issue of non-intended traffic.
%     \item Model TSV traffic (try to model it as bus).
%     \item Generate respective traffic traces using Gem5. 
%     \item Run simulation using trace based traffic.
% \end{itemize}

% \color{black}

% \begin{itemize}
%     \item Next steps: running gem5 and getting appropriate computation traces.
%     \item 
% \end{itemize}

\subsection{Discussion}

% \noindent
% \textbf{For one malicious routers :}

% In a $4 \times 4$ mesh NoC with deterministic XY routing, we analyze the probability that a single malicious router lies on the communication path between a randomly selected source-destination pair. Each source-destination pair has a unique XY path consisting of $|x_2 - x_1| + |y_2 - y_1| - 1$ intermediate routers. Excluding the source and destination, there are $14$ possible placements for a single malicious router. Using the known expected Manhattan distance between two random nodes in a $4 \times 4$ grid, which is $2.5$, the average number of intermediate routers becomes $1.5$. Therefore, the probability that the malicious router lies on a randomly chosen path is $\frac{1.5}{14} = \frac{3}{28} \approx 10.71\%$.

% For two disjoint path probability is zero. 

% \noindent
% \textbf{For two malicious routers:}

% Without any defense = 41.76\% chance of eavesdropping.

% With this defense = 4.41\% chance of eavesdropping.

Table~\ref{tab:eavesdrop} shows the eavesdropping probability under one and two-router threat models for both the baseline and our approach. The probabilities were computed by evaluating all possible permutations source-destination pairs and malicious router placements; a scenario was marked as an eavesdropping case if any malicious router lies on the packet path. In the case of a single malicious router, our method reduces the eavesdropping probability to 0\% for both mesh sizes (e.g., from 6.85\% to 0\% in $8 \times 8$). Even under the more pessimistic model with two malicious routers, our method limits eavesdropping probability to just 0.47\%. Compared to the AES-128 CTR approach, our AONT-based method achieves an approximate 79\% area reduction in a 64-node NoC, based on gate-equivalent (GE) estimates synthesized using a 28nm standard cell library. In contrast to AES, our approach also eliminates the overhead of key sharing and secure key storage, as it does not require a pre-shared key between the source and destination.

\begin{table}[ht]
\centering
\vspace{-0.05in}
\caption{Eavesdropping probabilities for different scenarios}
\vspace{-0.05in}
\label{tab:eavesdrop}
\begin{tabular}{|c|c|c|c|}
\hline
\textbf{NoC Size} & \textbf{Threat Model} & \textbf{Defense} & \textbf{\begin{tabular}[c]{@{}l@{}} Eavesdropping \\ Probability \end{tabular}} \\
\hline
\multirow{4}{*}{$4 \times 4$} 
& \multirow{2}{*}{1 malicious router} & No defense & $10.71\%$ \\
&                                     & Our approach   & $0\%$ \\
\cline{2-4}
& \multirow{2}{*}{2 malicious routers} & No defense & $41.76\%$ \\
&                                      & Our approach    & $4.41\%$ \\
\hline
\multirow{4}{*}{$8 \times 8$} 
& \multirow{2}{*}{1 malicious router} & No defense  & $6.85\%$ \\
&                                     & Our approach    & $0\%$ \\
\cline{2-4}
& \multirow{2}{*}{2 malicious routers} & No defense & $13.21\%$ \\
&                                      & Our approach   & $0.47\%$ \\
\hline
\end{tabular}
\vspace{-0.15in}
\end{table}
\section{conclusion}
\label{sec:conclusion}

This paper presented a lightweight confidentiality-preserving framework for secure on-chip communication in NoC architectures. By integrating a quasi-group-based AONT with disjoint multi-path routing, our approach ensures that intermediate routers cannot reconstruct the original message without access to all packet blocks. Experimental results on realistic workloads confirms that our approach provides confidentiality guarantees with 7.3 times reduction in performance overhead compared to AES, highlighting its effectiveness in defending against eavesdropping attacks.

\bibliographystyle{IEEEtran}
\bibliography{IEEEabrv,bibliography.bib}

\begin{thebibliography}{1}
\providecommand{\url}[1]{#1}
\csname url@samestyle\endcsname
\providecommand{\newblock}{\relax}
\providecommand{\bibinfo}[2]{#2}
\providecommand{\BIBentrySTDinterwordspacing}{\spaceskip=0pt\relax}
\providecommand{\BIBentryALTinterwordstretchfactor}{4}
\providecommand{\BIBentryALTinterwordspacing}{\spaceskip=\fontdimen2\font plus
\BIBentryALTinterwordstretchfactor\fontdimen3\font minus \fontdimen4\font\relax}
\providecommand{\BIBforeignlanguage}[2]{{%
\expandafter\ifx\csname l@#1\endcsname\relax
\typeout{** WARNING: IEEEtran.bst: No hyphenation pattern has been}%
\typeout{** loaded for the language `#1'. Using the pattern for}%
\typeout{** the default language instead.}%
\else
\language=\csname l@#1\endcsname
\fi
#2}}
\providecommand{\BIBdecl}{\relax}
\BIBdecl

\bibitem{weerasena2024security}
H.~Weerasena \emph{et~al.}, ``Security of electrical, optical, and wireless on-chip interconnects: A survey,'' \emph{ACM TODAES}, vol.~29, no.~2, pp. 1--41, 2024.

\bibitem{weerasena2021lightweight}
------, ``Lightweight encryption using chaffing and winnowing with all-or-nothing transform for network-on-chip architectures,'' in \emph{HOST}, 2021.

\bibitem{marnas2003all}
S.~I. Marnas \emph{et~al.}, ``All-or-nothing transforms using quasigroups,'' in \emph{Proc. 1st Balkan Conference in Informatics}, 2003, pp. 183--191.

\bibitem{gem5}
N.~Binkert \emph{et~al.}, ``The gem5 simulator,'' \emph{SIGARCH CA News}, 2011.

\bibitem{catania2016cycle}
V.~Catania \emph{et~al.}, ``Cycle-accurate network on chip simulation with noxim,'' \emph{ACM TOMACS}, vol.~27, no.~1, pp. 1--25, 2016.

\bibitem{woo1995splash}
S.~C. Woo \emph{et~al.}, ``The splash-2 programs: Characterization and methodological considerations,'' \emph{CA News}, vol.~23, no.~2, pp. 24--36, 1995.

\bibitem{bienia2008parsec}
C.~Bienia \emph{et~al.}, ``The parsec benchmark suite: Characterization and architectural implications,'' in \emph{PACT}, 2008, pp. 72--81.

\end{thebibliography}

% \newpage
% \import{sections/appendix}{points.tex}

\end{document}